\documentclass[article]{IEEEtran}
\usepackage{authblk}
\usepackage{amsfonts,amsmath,amsthm,amssymb,dsfont,bbm}
\usepackage{color}

\usepackage{balance} % for balancing columns on the final page
\usepackage[ruled,vlined, linesnumbered]{algorithm2e}
\usepackage{algorithmic}
\usepackage{enumitem}
\usepackage{graphicx}
\usepackage{tikz}
\usepackage{math}
\usepackage{float}
\usepackage{times}
\usepackage{caption}
\usepackage{subcaption}
\usepackage{pgfplots}
\usepackage{xcolor}
\usepackage{todonotes}
\newcommand{\fl}{f_{\lambda, L}}
\newcommand{\Ll}{L_\lambda}
\newcommand{\Lli}{L_{\lambda}^{(i)}}
\newcommand{\E}{\mathbb{E}}
\newcommand{\N}{\mathbb{N}}
\newcommand{\R}{\mathbb{R}}
\newcommand{\V}{\operatorname{Var}}

\newcommand{\M}{\mathcal{M}}
\newcommand{\Susc}{\mathcal{S}}
\newcommand{\Susct}{\tilde{\mathcal{S}}}
\newcommand{\I}{\mathcal{I}}
\newcommand{\It}{\Tilde{\mathcal{I}}}

\newcommand{\A}{\mathcal{A}}
\newcommand{\Ncal}{\mathcal{N}}
\newcommand{\Bcal}{\mathcal{B}}
\newcommand{\Prms}{\mathcal{P}}
\newcommand{\Map}{\mathbb{M}}
\newcommand{\Ti}{T^{(I)}}

\newcommand{\propspeed}{\mathcal{V}}
\newcommand{\rinf}{\mathcal{R}}

\newcommand{\Tti}{\tilde{T}^{(I)}}

\definecolor{commentcol}{RGB}{205,133,63}

\newtheorem{theorem}{Theorem}
\newtheorem{lemma}[theorem]{Lemma}
\newtheorem{corollary}[theorem]{Corollary}

\newtheorem{definition}[theorem]{Definition}

\providecommand{\keywords}[1]{\textbf{\textit{Index terms---}} #1}

\title{Agent-based Modeling and Simulation for Malware Spreading in D2D Networks}

\author[1]{Ziyad Benomar~\thanks{ziyad.benomar@orange.com}}
\author[1]{Chaima Ghribi\thanks{chaima.ghribi@orange.com}}
\author[1]{Elie Cali~\thanks{elie.cali@orange.com}}
\author[2]{Alexander Hinsen~\thanks{alexander.hinsen@wias-berlin.de}}
\author[2]{Benedikt Jahnel~\thanks{benedikt.jahnel@wias-berlin.de}}

\affil[1]{Orange Labs, Ch\^atillon, France}
\affil[2]{Weierstrass Institute, Berlin, Germany}

\begin{document}

\maketitle

\begin{abstract}
This paper presents a new multi-agent model for simulating malware propagation in device-to-device (D2D) 5G networks. This model allows to understand and analyze mobile malware-spreading dynamics in such highly dynamical networks.
Additionally, we present a theoretical study to validate and benchmark our proposed approach for some basic scenarios that are less complicated to model mathematically and also to highlight the key parameters of the model.
Our simulations identify critical thresholds for "no propagation" and for "maximum malware propagation" and make predictions on the malware-spread velocity as well as  device-infection rates. To the best of our knowledge, this paper is the first study applying agent-based simulations for malware propagation in D2D.
\end{abstract}


\section{Introduction}

\par D2D communications is one of the key emerging technologies for 5G networks and beyond. It enables a direct exchange of data between mobile devices, which extends coverage for devices lacking direct access to the cellular infrastructure and therefore
enhances the network capacity.
However, security issues are very challenging for D2D systems as malware can easily compromise mobile devices and propagate across the decentralized network. Compromised devices represent infection threats for all of their connected neighbors as they can, in their turn, propagate malware through susceptible devices and form an epidemic outbreak. This enables attackers to infect a larger population of devices and to launch cyber- and physical malicious attacks. Therefore, it is of great importance to have a good understanding of vulnerability and security issues, particularly of the malware propagation processes, in such networks and to be able to design optimal defense strategies.
\par Modeling malware propagation in D2D is challenging due to the complexity of such networks induced for example by topology or device mobility. In order to cope with this, D2D can be investigated and analyzed using analytical models (e.g., stochastic geometry, stochastic processes, etc.). Some of these approaches have been proposed to model malware spreading in D2D networks \cite{Zhang,Zhang1,Hinsen}. Nevertheless, classical simulation and analytical tools are often not suitable for capturing the global dynamics of complex systems.

 In this paper we propose to tackle the problem from the perspective of complex-systems science and present a new agent-based model (ABM) in order to analyze and understand malware propagation in D2D networks. For this, the agent-based simulation approach provides the possibility to simulate complex-systems dynamics and to test theories about local behaviors and their emergence. Unlike traditional techniques of simulation, based on mathematical or stochastic models, multi-agent simulation is more suitable for complex problem modeling and simulation. In fact, applying classical simulation and analytical tools, such as differential equations, to complex systems often produces undesired complications. Indeed, many challenges that arise in the traditional numerical modeling come from the fact that individual actions (activities that result in a modification of the system) and their impact on the dynamics of the system are often underrepresented. Usually, individual behaviors, i.e., decisions made at the individual or group level, cannot be incorporated into these simulations. On the other hand, in a multi-agent simulation, the model is not a set of equations as in mathematical models, but a set of entities. Here agents represent the set of all the simulated individuals, objects encode the set of all represented passive entities, and the environment is the topological space where agents and objects are located and which they can move in and act upon. 
 
Although agent-based simulations have been successfully used to model complex systems in different areas like biology, sociology, political science and economics, it is still insufficiently explored in the field of telecommunication networks, specifically for malware spreading in D2D. 
In this work, we aim to shed more light on whether such highly dynamical D2D networks can be treated as a complex system and whether complex-systems science can give insights on the emergent properties of malware propagation. 
The main contributions of this paper are as follows:
\begin{itemize}[noitemsep,nolistsep]
    \item We propose a new ABM for studying malware propagation in D2D 5G+ networks and we formally prove its correctness for predicting different agents status over the time. 
    \item We perform a theoretical study to estimate the critical values of the model's parameters and to identify the most important ones to consider for simulations. 
    \item We perform simulations to study and understand malware-spreading dynamics. Some critical thresholds     have been identified. Important aspects like malware infection rates and velocities have been also studied to understand how they will evolve as functions of the parameters. 
\end{itemize} 
The rest of the manuscript is organized as follows. Section~\ref{RL} reviews related work. Section~\ref{Model} describes the ABM for malware propagation in D2D networks. Section~\ref{MAsimulation} shows details of our multi-agent simulation implementation. Section~\ref{TS} presents a theoretical study for the problem in some specific scenarios. Section~\ref{PE} shows simulation results followed by conclusions in Section~\ref{conc}.
\vspace{-0.1cm}
%%%%%%%%%%%%%%%%%%%%%%%%%%%%%%%%%%%%%%%%%%%%%%%%%%%%%%%%%%%%%%%%%%%%%%%%
\section{Related work}
\label{RL}
ABMs are effective and robust tools in simulating complex and dynamic phenomena like epidemic spreading. These models have been used primarily in epidemiological studies of infectious diseases and have recently gained a great importance also in the epidemiological modeling as can be seen from the vast literature in the context of the COVID-19 pandemic, see for example \cite{muller2021predicting,blakely2021determining,maziarz2020agent,drogoul2020designing}.

However, ABMs are still in their infancy with regard to telecommunication networks. Some ABMs have been proposed in the literature for IoT networks. Authors in \cite{ABMiot,Butt,Perez} proposed ABMs for analyzing IoT systems. Other applications of ABMs to telecommunication networks are proposed in \cite{Niazi1} and \cite{Niazi2}, where authors analyzed the effectiveness of ABMs to understand self-organization in peer-to peer and ad-hoc networks. These studies provide further motivation to our investigation on applying ABMs for studying malware spreading dynamics in D2D 5G networks.

Let us again mention that conventionally D2D systems are modeled using analytical methods (e.g., stochastic geometry) which have proven to be powerful tools for modeling spatial device and road systems. 
In this context, the authors in \cite{Zhang} and \cite{Zhang1} present a framework for the modeling and understanding of malware spread in D2D with mobile devices and study some strategies of both defenders and  attackers. The proposed model is based on an analytical approach and does not consider urban environments. In view of this, a standard SIR model is presented in \cite{ChaseEscape}, to study malware propagation in D2D considering urban environments but mobility was not taken into account. Even though the obtained results were promising, some questions remained open regarding the convergence of the malware propagation speed, the shape theorem of the infection and the critical thresholds. This mainly comes from the fact that the dynamics of the system were insufficiently captured. 
\section{System model}
\label{Model}
This section gives a detailed description of the D2D malware propagation model in urban environments. In this ABM description, devices are represented as reactive agents that move in the environment and have a variety of capabilities like neighborhood discovery and malware propagation. In short terms, the system has the following composition. We consider an urban environment. At initial time, devices are placed randomly on the streets (we make the simplifying assumption that devices that are situated in buildings are not to be taken into account : this can be justified by the high frequencies used in 5G). The devices move independently and randomly at a constant speed.
Moreover, two devices can communicate directly with each other if they are close enough and on the same street. Let us note that this approach takes shadowing into account, but not interference.
At time zero, a virus is introduced carried by a device near to the center of the city. The virus can now propagate from one device to another if they can communicate for a sufficiently long time that represents the discovery time plus the transmission time. 
\subsection{Street systems and devices}
\label{pvt}
We consider our urban street environment $E$ as a two-dimensional planar Poisson--Voronoi tessellation (PVT, see \cite{Chiu_2013}) induced by a homogeneous Poisson point process (PPP) $X_{E}$ of positive intensity $\lambda$. The PVT \textcolor{black}{ is a one-parameter segment process that has been shown to be a good fit for the street systems of European cities (see \cite{Gloaguen_2006Fitting,Gloaguen_2009,Gloaguen_2017}). 
It has been widely used to model different urban environments as random tessellations, since it allows to go beyond specific urban topologies.} We will denote by $S$ the set of edges of $E$ (representing the streets).
The devices are placed on $S$ as a linear PPP of intensity $\theta$, thus forming a Cox point process (CPP) on the plane with random intensity measure $\Lambda(B) = \theta |S \cap B |$ for every measurable $B \in \R^2$. Here $|S \cap B |$ stands for the total length of $S$ in the area $B$.
%%%%%
\subsection{ABM for malware propagation in D2D}\label{sec:ABM}
We note first that the environment is modeled as an undirected graph, relying on some stochastic-geometry concepts, as described in Section~\ref{pvt}. 
Then, we define our malware-propagation system in D2D as a finite number of agents, states, actions and rules,
\begin{center}
$\rm MAS := \{\A, St, Act, R, T \}$.
\end{center}
More precisely, we consider a set of $n$ {\em agents} $\A = \{\,a_i \colon i \in [1,n]\,\}$ corresponding to devices and a state space ${\rm St} =  \{ \text{susceptible}, \text{infected}\}$. Further, ${\rm Act} = \{ \text{"move", "discover", "connect", "infect"} \}$ denotes the set of {\em actions} that each agent can perform according to its state. $\rm R$ represents the set of the {\em behavioral rule base}.
Time $\rm T$ is assumed to be divided in time units called {\em slots}, where each slot $k$ is represented by a positive integer.

Initially, agents are distributed on the edges of $E$ (i.e., streets of the city) as described in Section \ref{pvt}. One agent of type \textit{infected} is introduced around the center of the map. Then, formally, each agent $a_i$ is defined at each time slot by a tuple 
$$\M_{i,k} := \{ X_{i,k}, V_{i,k}, N_{i,k}, {\rm Act}_{i,k}, \xi_{i,k}, \Ti_{i,k} \}.$$
Here, $X_{i,k}$ specifies the agent's {\em location} in terms of coordinates at time $kdt$, $V_{i, k} = v$ represents the agent's moving {\em speed} and $N_{i,k}$ the {\em knowledge base}, representing what each agent $a_i$ knows about its neighboring agents and the environment at time slot $k$. Further, $\xi_{i,k} \in \rm St$ represents the state of agent $a_i$ and ${\rm Act}_{i,k}$ is the set of actions that can be performed by $a_i$.
Finally, $\Ti_{i,k}$ represents the {\em first time} when $a_i$ becomes infected. It will be updated during the simulation depending on the agent's interactions. 
$\Ti_{i,0}$ is set to $+ \infty$ for initially susceptible agents and $0$ for the infected one. The state of $a_i \in \A$ at $kdt$ for $k\geq 1$ is given by
\[
\xi_{i,k} := 
\left\{
    \begin{array}{ll}
        \text{susceptible} &\text{if } kdt < \Ti_{i,k-1},\\
        \text{infected} &\text{if } kdt \geq \Ti_{i,k-1}.\\
    \end{array}
\right.
\]
In particular, the state of $a_i$ at a step $k$ of the simulation is computed using the variables $(\Ti_{i,k-1})_{a_i \in \A}$ from the previous step. This formula also implies that the states of the agents will not change between the steps $0$ and $1$. It will be indeed the case since we will consider a time step $dt$ smaller than $\rho$. (See Section~\ref{sec:correctness}).
\subsection{Agent behavior and states}
Let us describe the three different behaviors of agents: mobility, communication and infection.

\subsubsection{Mobility behavior}
\label{sec:mob}
Devices move at the same constant speed $v$\textcolor{black}{, starting from a base position, and} repeating indefinitely the following  {\em street-adapted random-waypoint model}: 
\begin{itemize}
    \item Each device independently picks a destination on the street. For this we sample a random point $P$ in the plane using a Gaussian distribution centered on the device $X$, and with a standard deviation equal to $\sigma_X = (15 {\rm min})\times v$. The {\em destination} we take for $X$ is then the closest point of $P$ in $E$. This choice of $\sigma_X$ shows that devices will go to destinations that they can reach in an average time of $15 {\rm min}$ if they take a straight path.
%%%%%
    \item Devices move to their destinations following the shortest path along the streets.
    \item Once arrived, devices go back to their starting position following the shortest path along the streets \textcolor{black}{(anchored movement)}.
\end{itemize}

\subsubsection{Communication behavior}
\label{sec:combehavior}
In order to exchange messages, two communicating devices/agents must obey the following rules:
\begin{itemize}
\item (RAD): The Euclidean distance between the two devices is less than a given constant threshold $r$.
\item (LOS): The two devices are on the same street.
\end{itemize}
The first rule supposes that the emission power of the devices is a constant and that we do not take into account interference.
The second rule means that the signal cannot go through the buildings and that reflections and diffractions are not taken into account. In symbols, for $X_i(t)$ the position of device $a_i$ at time $t$ and
$\Ncal(a_i, a_j) := \{t \geq 0 \colon \|X_i(t) - X_j(t)\| < r \hbox{ and } \exists s \in S \hbox{ such that } (X_i(t),X_j(t)) \in s \}$,
we have that 
$a_i$ and $a_j$ are connected at time $t$ if and only if $t \in \Ncal(a_i, a_j)$.

\subsubsection{Infection behavior}
We will follow a standard SI compartmental model, very similar to SIR which is a classical approach in epidemiology often used within the framework of differential equations. However, unlike the latter, in a D2D context, users are constrained to be positioned on streets and are mobile, two aspects that are usually not represented in epidemiological studies. The SI model is formulated by first partitioning devices into two distinct categories called susceptible (S) and infected (I). At time zero, only one device will be in the infected state, while a CPP $X_S$ with intensity $\theta$ will define the susceptible devices, independent of the former one given the PVT tessellation. When an infected device is connected to a susceptible device for a time longer than a given threshold $\rho$, the susceptible device will become infected.
More precisely, if the device $a_i$ is infected at time $t$ and if $[t, t+\rho] \in \Ncal(a_i, a_j)$, then $a_j$ is infected at $(t+\rho)$.
\subsubsection{Agent states}
Agent states specify what state an agent is in. Agent-state transitions are driven by the rule base $\rm R$
that implements the reactive behavior of agents. It allows to select actions to take for agent $a_i$ depending on its current local state $\xi_{i,k}$ and its knowledge base $N_{i,k}$.
More specifically, we write ${\rm R}=\{ \Theta\}$ where $\Theta( \xi_{i,k}, N_{i,k})$ are the {\em active rules} that map the set of states and observations to actions for reactive tasks
\begin{center}
 $\Theta : ( \xi_{i,k}, N_{i,k}) \longrightarrow {\rm Act}_{i,k}$.
\end{center}
Let $T^{(C)}_{i,j}$ be the connection duration between agents $a_{i}$ and $a_{j}$ and $\rho$ be the needed time for the virus transmission from one agent to another. Then the principal rule-based function is described as follows. 
\begin{itemize}
    \item Malware infection rule: If agent $a_{i}$ is infected, agent $a_{j}$ is susceptible ($\xi_{i,k} = \text{infected}$, $\xi_{j,k} = \text{susceptible}$ ) and $a_{i}$ was connected to $a_{j}$ for a time longer than the infection threshold ($T^{(C)}_{i,j} \geq \rho$), then the state of agent $a_{j}$ will be transited from susceptible to infected (the action infect will be activated),    \begin{center}
    $\Theta_{I} : ( \xi_{i,k}, N_{i,k})     \longrightarrow \textbf{Infect}$.   
    \end{center}
\end{itemize}
A more detailed description of the algorithm associated to malware infection will be given in Section \ref{MAsimulation}.

%%%%%%%%%%%%%%%%%%%%%%%%%%%%%%%%%%%%%%%%%%%%%%%%%%%%%%%%%%%%%%%%%%%%%%%%
\section{Agent-based simulation}
\label{MAsimulation}
In this section we present more details on the implementation of our multi-agent simulation tool. 
Let us denote by 
\[ \Prms := \{ dt, \rho, r, \lambda, \theta, v \}\]
the set of key model parameters where $dt$ represents the elapsed time in each step, $\rho$ and $r$ represent respectively connection time needed for virus transmission and communication radius of agents. $\lambda$ is the intensity of Voronoi seeds (seed/$\text{km}^2$), $\theta$ is the intensity of susceptible agents (agent/km) and $v$ denotes agents speed (km/h). 
Other parameters such as the dimensions $(H_1, H_2)$ of the map can be added to this list, but we will not focus on these in our study. For the most part of the manuscript, we give the same speed to all the agents in order to keep a restraint number of parameters. However, we can easily have a more general model where the speeds of the agents are distributed following some probability law. Each agent could have for example a speed taken uniformly at random in some interval $[v_1, v_2]$.\\
Our simulation is done over steps, each step corresponds to a time instant $kdt$. In the following we will denote by $\M_k$ the model at step $k$. 
It represents the map, the agents and all their attributes (coordinates, states, etc.) at step $k$. 
In the simulation, we first generate a random map, then the agents, and after that we run the function Step($\M_k$), that updates the variables of the model, taking it from a step $k$ to the next step $k+1$, for a number $k_{\max}$ of iterations. Algorithm \ref{algo:main} describes the entry function of the simulation.
\begin{algorithm}[h]
   \SetKwInOut{Input}{Input}
   \SetKwInOut{Output}{Output}
   \caption{Main($\Prms,k_{\max} $): The main function describing the simulation}\label{algo:main}
   \Input{The set of parameters $\Prms$ and the maximum number of steps $k_{\max}$}
   \Output{The state of a randomly generated model at time $k_{\max}dt$}
   $\Map \gets $ GenerateMap($\lambda$)\;
   $\A \gets$ GenerateAgents($\theta$)\;
   $\M_0 \gets (\Prms, \Map, \A, (X_{i,0})_i, (\Ti_{i,0})_i)$\;
   \For{$k \in \{1, \ldots ,k_{\max}\}$}{
        $\M_k \gets \operatorname{Step}(\M_{k-1})$\;
   }
   \KwRet{$\M_{k_{\max}}$}
\end{algorithm}
The function GenerateMap($\lambda$) returns a random PVT with parameter $\lambda$, whereas the function Generate($\theta,v$) returns the set of agents $\A := \A_S \cup \{ a_{i_0} \}$, where $\A_S$ is the set of initially susceptible agents distributed on $\Map$ using homogeneous PPP with parameters $\theta$. $a_{i_0}$ is the initially infected agent, placed near the center of the map.

\subsection{Discrete-time approximations}
Recall that our simulations are done over steps, where each step $k$ corresponds to a time instant $k dt$. 
A difficulty lies in the correct updating of the states of the agents. From step $k$ to $k+1$, each agent moves independently as described in Section \ref{sec:mob}, which means that we can access the positions of the agents at times $(kdt)_{k\in \N}$ knowing their velocities and the edges they have been through, but it is complicated to know all the interactions they had given only this information.
To overcome this, we first impose the constraint $dt < \rho$. This guarantees that, by only observing the positions of the agents at discrete times with a step $dt$, we will not miss any two devices that connect for a duration longer than $\rho$, see Section~\ref{sec:correctness} for more details.
Let $k \in \N$, and let us assume that $a_i,a_j$ are connected to each other at $kdt$. We will treat the general case where they can have different speeds $v_i$ and $v_j$, and we will compute the duration of the connection using their movement equations. 
Let us denote by $t^{(\text{in})}_{i,s}$ (respectively $t^{(\text{out})}_{i,s}$) the time when $a_i$ gets in (respectively out) of the street $s$. These can easily be computed knowing $X_i$ and the length $L(s)$ of the street $s$. Since $s$ has two different directions, we need to consider their velocities $\textbf{v}_i,\textbf{v}_j$. Let $P_1,P_2$ be the positions of the two extremities of the street $s$, let $\textbf{e} := (P_2 - P_1)/\| P_2 - P_1 \|$ (we can take $-\textbf{e}$ instead), and $\nu_i,\nu_j$ be such that $\textbf{v}_i = \nu_i\textbf{e}, \textbf{v}_j = \nu_j\textbf{e}$. 
We recall that the absolute speed $v_i$ of $a_i$ obeys  $v_i = \| \textbf{v}_i \| = \pm \nu_i$, the same holds for $a_j$.
Finally, let us also define the coordinates of $a_i,a_j$ on the street $s$ by $d_{i,k} := (X_{i,k} - P_1)\cdot \textbf{e}$ and $d_{j,k} := (X_{j,k} - P_1)\cdot \textbf{e}$. Then we have the following result that we present without proof.

\begin{lemma}\label{lem:cnct_interval}
If $a_i,a_j$ are connected at time $kdt$ and if $\nu_i \neq \nu_j$, then they are connected during all the time interval $[t^{(C,i)}_{i,j},t^{(C,f)}_{i,j}]$, where
\begin{align*}
    t^{(C,i)}_{i,j} &:=
    \max\{ kdt - \dfrac{d_{i,k} - d_{j,k}}{\nu_i - \nu_j} - \dfrac{r}{|\nu_i - \nu_j |} ,t^{(\text{in})}_{i,s}, t^{(\text{in})}_{j,s} \},\\
    t^{(C,f)}_{i,j} &:=
    \min\{ kdt - \dfrac{d_{i,k} - d_{j,k}}{\nu_i - \nu_j} + \dfrac{r}{|\nu_i - \nu_j |} ,t^{(\text{out})}_{i,s}, t^{(\text{out})}_{j,s} \}.
\end{align*}
Moreover, if $\nu_i = \nu_j$, then 
\[
t^{(C,i)}_{i,j} =
\max\{t^{(\text{in})}_{i,s}, t^{(\text{in})}_{j,s} \} \text{ and } t^{(C,f)}_{i,j} =
\min\{t^{(\text{out})}_{i,s}, t^{(\text{out})}_{j,s} \}.
\]
The connection duration of $a_i,a_j$ is then $T^{(C)}_{i,j} := t^{(C,f)}_{i,j} - t^{(C,i)}_{i,j}$.
\end{lemma}
\textcolor{black}{In words, two agents on the same street can have different speeds and move either in the same or in opposite directions. Recall that the connection-time interval is the set of all times 
such that the distance of the two agents is less than $r$.}

We saw in Section \ref{sec:ABM}, that agents states will be determined by the variable $\Ti_{i,k-1}$ at each step $k \geq 1$. We call $\Susc_k,\I_k$ the sets of susceptible and infected agents. Let ConnectionInterval($a_i,a_j,k$) be a function computing $t^{(C,i)}_{i,j}, t^{(C,f)}_{i,j}$ as in Lemma \ref{lem:cnct_interval} knowing that $a_i,a_j$ are connected at $kdt$, and let GetNeighbors($a_i$) be a function returning the set of neighbors of $a_i$ defined as: $N_k(a_i) := \{ a_j \in \A  \colon \|X_{i,k} - X_{j,k}\| \leq r$ $\text{ and } a_i,a_j \text{ on the same street}\}$. Finding the neighbors of all the agents would normally require a $O(n^2)$ time complexity, but since only agents on the same street can connect to each other, we can considerably reduce this complexity by searching neighbors of each agent only among those that are on the same street. From here, we can write Algorithm \ref{algo:infect} that updates the values $\Ti_{j,k}$ for the neighbors of an infected agent $a_i$.

\begin{algorithm}
   \SetKwInOut{Input}{Input}
   \SetKwInOut{Output}{Output}
   \caption{InfectNeighbors($a_i$)}\label{algo:infect}
   \Input{An infected agent $a_i$}
   \Output{Updates $\Ti_{j,k}$ for all susceptible neighbors of $a_i$}
   $N^{(S)}_k(a_i) \gets \operatorname{GetNeighbors}(a_i) \cap \Susc_k$\;
   \For{$a_j \in N^{(S)}_k(a_i)$}{
        $t^{(C,i)}_{i,j},t^{(C,f)}_{i,j} \gets \operatorname{ConnectionInterval}(a_i, a_j,k)$\;
        $t^{(C,i)}_{i,j} \gets \max\{ t^{(C,i)}_{i,j}, \Ti_{i,k} \}$\; \label{algoline:ti=min}
        \If{$T^{(C)}_{i,j} := t^{(C,f)}_{i,j} - t^{(C,i)}_{i,j} \geq \rho $}{
            $\Ti_{j,k} \gets \min\{ \Ti_{i,k}, t^{(C,i)}_{i,j} + \rho \}$\; \label{algoline:Ti=min}
        }
   }
\end{algorithm}
Line~\ref{algoline:ti=min} makes sure that we only compute the time when the agents are connected and $a_i$ is infected. Note that in Line~\ref{algoline:Ti=min}, we cannot set the value of $\Ti_{j,k}$ simply to $t^{(C,i)}_{i,j} + \rho$ as agent $a_j$ might be connected to several infected agents, and it will become infected as soon as it stays connected to one of them for longer than $\rho$. Finally, we can write the core function of our simulation, that is Algorithm \ref{algo:step}.
\begin{algorithm}[h!]
   \SetKwInOut{Input}{Input}
   \SetKwInOut{Output}{Output}
   \caption{The Step Function}\label{algo:step}
   \Input{The model $\M_{k-1}$ at step $k-1$}
   \Output{The model $\M_{k}$ at step $k$}
   
   $\Susc_k, \I_k \gets$ The sets of susceptible and infected agents\;
   
   \For{$a_i \in \A$ }{
        $X_{i,k} \gets \operatorname{Move}(a_i,V_i,X_{i,k-1}, dt)$; \textcolor{commentcol}{//Update the positions} \\
        $\Ti_{i,k} \gets \Ti_{i,k-1}$; \textcolor{commentcol}{//Initialisation} \\
   }
   \For{$a_i \in \I_k$}{
        $\operatorname{InfectNeighbors}(a_i)$; \textcolor{commentcol}{//Update the variables $\Ti_{j,k}$}\\
   }
   $\M_{k} \gets (\Prms, \Map, \A, (X_{i,k})_i, (\Ti_{i,k})_i $\;
   \KwRet{$\M_k$}\;
\end{algorithm}
\vspace{-0.2cm}
\subsection{Equivalence of discrete and continuous time}\label{sec:correctness}
We denote by $\xi_i(t)$ the state of agent $a_i$ at {\em continuous time} $t$ for any $a_i \in \A$. On the other hand, for each $k\in \N$ we denote as before by $\xi_{i,k}$ the state of $a_i$ at {\em discrete time} $kdt$ as predicted by our ABM. The following theorem states that for sufficiently small time slots, at the discrete time points, our model is equivalent to its continuous-time version and is then theoretically proven to be correct. 
\begin{theorem}\label{th:correctness}
If $dt < \rho$, then we have 
\[
\forall a_i\in \A, \forall k \in \N, \qquad \xi_{i,k} = \xi_i(kdt).
\]
\end{theorem}
\textcolor{black}{In words, Theorem~\ref{th:correctness} guarantees that, by discretizing, we do not miss infection events and the introduced time differences do not induce errors in the discretized model.}
Let us first define the first continuous time when $a_j \in \A$ is infected, i.e.,  $\Tti_j := \inf\{t \geq 0 \colon \xi_j(t) = \text{infected} \}$. Regarding our malware propagation rules, we can write
\begin{equation}\label{eq:Tti}
\Tti_j = 
\inf\limits_{a_i \neq a_j} \;
\inf\limits_{ t \geq \Tti_i}
\{
t + \rho \colon
[t, t+\rho] \subset \Ncal(a_i, a_j)
\},
\end{equation}
where $\Ncal(a_i, a_j)$ is as defined in Section~\ref{sec:combehavior}. Let us also denote $\Susc_k := \{a_i \colon kdt < \Ti_{i,k-1} \}$, $\Susct_k := \{a_i \colon kdt < \Tti_{i} \}$, $\I_k := \{a_i \colon kdt \geq \Ti_{i,k-1} \}$ and $\It_k := \{a_i \colon kdt \geq \Tti_{i} \}$. Finally, for convenience, let $\Ti_{i,-1} := \Ti_{i,0}$ for all $a_i \in \A$. We have the following lemma.
\begin{lemma}\label{lem:correctness}
If $dt < \rho$, then for any $k \in \N$, assertion $\Bcal_k$ is true
\[
(\Bcal_k) \; : \qquad \forall a_j \in \A, \;
\left\{
    \begin{array}{l}
        \Tti_j \leq \Ti_{j,k-1},\\
        \Tti_j \leq kdt \implies \Tti_j = \Ti_{j,k-1}.
    \end{array}
\right.
\]
\end{lemma}
Note that, if $\Bcal_k$ is verified for some $k \in \N$, then $\Susct_k \subset \Susc_k$ and $\It_k \subset \I_k$. But since $\Susct \cup \It_k = \Susc_k \cup \I_k$, this means that $\Susct_k = \Susc_k$ and $\It_k = \I_k$ and thus Theorem~\ref{th:correctness} is proved.

\begin{proof}
For $k=0$ the assertion is true by definition of $(\Ti_{i,-1})_{a_i \in \A}$. Let $k \geq 1$, assume that $\Bcal_k$ is true and let $a_j \in \A$. If $\Ti_{j,k} = \Ti_{j,k-1}$ then directly $\Tti_j \leq \Ti_{j,k}$. Otherwise $\Ti_{j,k}$ was updated during step $k$, i.e., there exists an agent $a_i \in \I_k$ for which InfectNeighbor($a_i$) was called and such that $a_j \in N_{i,k}^{(S)}$ and $t_2 - t_1 \geq \rho$ with $t_1 = \max\{ t^{(C,i)}_{i,j}, \Ti_{i,k-1} \}, \, t_2 = t^{(C,f)}_{i,j}$. This implies that $[t_1, t_1 + \rho] \subset [t^{(C,i)}_{i,j}, t^{(C,f)}_{i,j}] \subset \Ncal(a_i,a_j)$, and since $a_i \in \I_k$, we have by the induction hypothesis that $\Tti_i = \Ti_{i,k-1}$ and thus $t_1 \geq \Tti_i$. Thus, using Equation (\ref{eq:Tti}), we have that $\Tti_j \leq \Ti_{j,k}$. For the second part of the assertion, let us assume that $\Tti_j \leq (k+1)dt$. If $\Tti_j \leq kdt$ then $\Ti_{j,k} \leq \Ti_{j,k-1} = \Tti_j$ (induction hypothesis), and we proved that $\Tti_j \leq \Ti_{j,k}$ and therefore $\Tti_j = \Ti_{j,k}$. Otherwise $kdt < \Tti_j \leq (k+1)dt$, this implies that $a_j \in \Susct_k$ and there exists $a_i \in \A$ such that $[\tilde{t}, \tilde{t} + \rho] \subset \Ncal(a_i, a_j)$ and $\tilde{t} \geq \Tti_i$ with $\tilde{t} := \Tti_j - \rho$. Given that $dt < \rho$ we have
\[
\Tti_i 
\leq \tilde{t} 
= \Tti_j - \rho_I
\leq (k+1)dt - \rho_I
< kdt 
< \Tti_j = \tilde{t} + \rho_I,
\]
and this implies that $a_i \in \It_k$ and $kdt \in [\tilde{t}, \tilde{t} + \rho] \subset \Ncal(a_i, a_j)$. Thus InfectNeighbors is called on $a_i$ at step $k$ and $a_j$ is among the visited agents during this call (neighbors of $a_i$). $\Tti_{j,k}$ will then be updated and its final value will be at most $\tilde{t} + \rho = \Tti_i$. With the inequality $\Tti_j \leq \Ti_{j,k}$ that we already proved, we deduce that $\Tti_j = \Ti_{j,k}$.
\end{proof}
Finally, for any $k \in \N$, we have by Lemma~\ref{lem:correctness} that  $\Susct_k = \Susc_k$ and $\It_k = \I_k$. This means that the states of the agents predicted by the simulator correspond to their real states.

\section{Mean-field version}
\label{TS}
The model that we presented so far is very rich with many parameters. It is therefore difficult to run simulations varying all these parameters and see how each of them influences the propagation of the virus. So, in order to better choose the values we will assign to them, in this section, we present a theoretical study on a simplified model to identify critical relationships between parameters and values that will lead to drastic changes in the system's evolution.
\textcolor{black}{Let us highlight that we consider a different model that does not arise as a limiting object. It is mainly introduced in order to sharpen the intuition for threshold values of important parameters.}

As in the first model, we start with a single infected agent $a_{i_0}$, and we will take interest in the time of the first virus transmission, which we will denote by $\tau$ in the following. Let us stress that the simplified model that we present here is used only as a mathematical model. All the simulations results in Section~\ref{sec:simres} are based on the original model and not this simplified one.

We consider the following mean-field approximation of our spatial model. Instead of considering $a_{i_0}$ to be moving on a PVT, we will consider that it moves on a succession of streets $s_0, s_1, \ldots$, each having a length $\Lli$ that is a random variable with density $\fl$, where $\fl$ is the density function of the edges lengths in a PVT having a seeds intensity equal to $\lambda$ (see Section~\ref{pvt}). We will assume that, when $a_{i_0}$ enters a street, other agents are distributed on it as an homogeneous PPP with parameter $\theta$, and that they can move in any of the two possible directions.
What we mainly lose in this simplified model is the dependence between the lengths of the successive streets visited by $a_{i_0}$. 

For each street $s_i$ visited by $a_{i_0}$, let $C_i$ be the number of agents that $a_{i_0}$ infects while being on $s_i$. Let $p := \Pr[C_i \geq 1]$ denote the probability that $a_{i_0}$ infects at least some agent on $s_i$ ($p$ is independent of $i$). 
\[
\tau := \inf \{t\geq 0 \colon \exists j\neq i_0 \text{ such that } \xi_{j}(t) = \text{infected} \}.
\]
Then, we have the following main results.
\begin{theorem}\label{th:lbexp}
If $\tau$ is the first time when $a_{i_0}$ infects another agent, then 
\[
\frac{2}{3\sqrt{\lambda} v}(1/p-1) \leq \E [\tau] \leq \frac{2}{3\sqrt{\lambda} v}\cdot 1/p.
\]
\end{theorem}
\begin{theorem}\label{th:lbprob}
There exists a positive constant $\Tilde{C}$ such that if $p$ is sufficiently small, then for $t_0 = 1/(3\sqrt{p\lambda}v)$ we have
\[
    \Pr [ \tau \geq t_0 ] 
    \geq 1 - \Tilde{C}p^{1/4}.
\]
\end{theorem}
These theorems indicate that, if the probability of infecting another agent on a single street $s_i$ is low, then the waiting time before the virus transmission is very large, and therefore the virus propagation is weak. In terms of the asymptotic behavior of the system, we can state that, when $p = o(1)$, then $\E[\tau] = \Omega(1/(\sqrt{\lambda}vp)$ and for $t_0 = 1/(3\sqrt{p\lambda}v)$ we have $\Pr[\tau \geq t_0] = 1 - O(p^{1/4})$.

The proofs rely on results for typical edge length in PVT and Berry--Esseen inequalities. 
Let us start by presenting a first lemma on the edges-lengths distribution when $\lambda = 1$, as described in \cite{Vorplane}.
\begin{lemma}\label{lem:f1}
In a random planar PVT, if we choose a random edge, then its length $L$ is a random variable having a distribution $f_L$ satisfying
\begin{enumerate}
    \item $f_L(0) = 2/\pi$, when $l$ is large enough: $f_L(l) \sim \tfrac{\pi^2}{3\sqrt{2}}l^2e^{-\frac{\pi}{2} l^2}$,
    \item if $L$ is a random variable with density function $f_L$, then $L$ has an $n$-th moment for any positive integer $n$ and
    \item $\E[L] = 2/3$,  $\sigma_L^2 := \V[L] \approx 0.1856$.
\end{enumerate}
\end{lemma}
From this we deduce the result for any positive $\lambda$.
\begin{lemma}\label{lem:normalize_f1}
If $P_\lambda$ is a PVT generated with an intensity of seeds $\lambda > 0$, then the edges length in $P_\lambda$ will have a density function $\fl$ given by
\[
\fl (l) := \sqrt{\lambda}f_L(\sqrt{\lambda}l),\qquad \forall l \geq 0.
\]
\end{lemma}
Then we have the following statement.
\begin{corollary}\label{cor:moments}
If $\Ll$ is a random variable with density $\fl$, then for any positive integer $n$, $\Ll$ has a $n$-th moment given by
\[
\E [\Ll^n] = \lambda^{-n/2}E[L_1^n].
\]
\end{corollary}
Finally, since $f_L$ is a rapidly decreasing function when $\ell$ is large, we have the following probability estimate.
\begin{lemma}\label{lem:x>l_0}
There exists $l_0 > 0$ such that for any  $x \geq l_0/\sqrt{\lambda} $
\[
\Pr [\Ll \geq x] \leq \exp (-\lambda x^2).
\]
\end{lemma}
Next, the following theorem is a corollary of the Berry--Esseen's inequality~\cite{Berry, Esseen} applied to random variables $(X_i - \mu)$ and using the trivial relation $\Pr[Y_n > x] = 1 - \Pr[Y_n \leq x]$ for $Y_n := \frac{1}{n}\sum\limits_{i=0}^{n-1} (X_i - \mu)$.
\begin{theorem}\label{cor:berry}
There exists a constant $C$ such that if $X_0, X_1, \ldots$ are i.i.d.~random variables with $\E[|X_0|] = \mu < +\infty$, $\V[X_0] = \sigma^2 >0$ and $\E[|X_0 - \mu|^3] = m < +\infty$, then for any $n \in \N$ and $x \in \R$
\[
\Pr \left[ \sum\limits_{i=0}^{n-1} X_i > x \right] \geq 1 - \Phi\left( \left(\frac{x}{n} - \mu \right)\frac{\sqrt{n}}{\sigma} \right) - \frac{C m}{\sigma^3\sqrt{n}}.
\]
Here $\Phi$ is the cumulative distribution function of the standard normal distribution.
\end{theorem}
We are now in the position to prove our main theorems. 
\begin{proof}[Proof of Theorem \ref{th:lbexp}]
We only need to observe that $\tau_0 + \ldots+ \tau_{m-1} \leq \tau \leq \tau_0 + \ldots +\tau_{m}$, where $m$ is the index of the first street such that  $C_m \geq 1$ and $\tau_i := L_\lambda^{(i)}/v$ is the time spent by $a_{i_0}$ on the street $i$. Using the law of total expectation, we deduce that $\E[m]\E[L_\lambda]/v \leq \E[\tau] \leq (\E[m] + 1)\E[L_\lambda]/v$, and the result is obtained by computing the two expectations $\E[m]$ and $\E[L_\lambda]$.
\end{proof}

\begin{proof}[Proof of Theorem \ref{th:lbprob}]
To prove this theorem, we first need to observe that for any $\tilde{m} \in \N$ we have
$\Pr [ \tau \geq t_0 ] 
\geq \Pr [ \tau_0 + \ldots +\tau_{m-1} \geq t_0 ] 
\geq \Pr [ \tau_0 + \ldots +\tau_{\tilde{m}-1} \geq t_0 ]\Pr [m \geq \tilde{m}] $ (the second inequality is true because all the $\tau_i$ are non-negative). In particular for $\tilde{m} =  \lceil 1/\sqrt{p} \rceil \leq 1/\sqrt{p} + 1$, Bernoulli's inequality gives that
\[
\Pr [m \geq \tilde{m}] = (1-p)^{\tilde{m}} 
\geq 1-\tilde{m}p \geq 1 - \sqrt{p} - p
\geq  1 - 2\sqrt{p}.
\]
On the other hand, if $p\leq (6\sigma_L)^{-4}$, using Theorem \ref{cor:berry} and the inequality $\Phi (-x) \leq \exp(-x)/\sqrt{2\pi}$, which is true for any $x\geq 2$, we find a constant $C_1$ verifying 
\[
\Pr [ \tau_0 + \ldots +\tau_{\tilde{m}-1} \geq t_0 ] \geq 1 - C_1 p^{1/4}.
\]
Finally, when $p$ is small enough, we deduce using again Bernoulli's inequality that
\[
\Pr [ \tau \geq t_0 ] 
\geq (1 - 2\sqrt{p})(1 - C_1 p^{1/4})
\geq 1 - (2 + C_1)p^{1/4}.
\]
This finishes the proof.
\end{proof}
We will now apply the previous theorems to show that the virus propagation is slow in any of the following cases.
\begin{itemize}
    \item The transmission time of the virus is very large compared to the expected time spent by agents on streets : $\sqrt{\lambda}\rho v \gg 1$
    \item The number of agents reachable within the communication radius is very small: $\theta r \ll 1$
    \item The number of agents on each street is very small: $\theta/\sqrt{\lambda} \ll 1$.
\end{itemize}

\begin{corollary}\label{cor:lrv>>1}
If $\sqrt{\lambda}\rho v \geq l_0$, then 
$
\E [ \tau ] \geq \dfrac{2}{3\sqrt{\lambda} v} (e^{\lambda \rho^2 v^2} - 1).
$\end{corollary}
\begin{proof}
We have the implications
$
C_0 \geq 1 \Rightarrow \tau_0 \geq \rho \Rightarrow L_{\lambda}^{(0)} \geq \rho v
$.
Hence, using Lemma \ref{lem:x>l_0},  
$
p = \Pr [ C_0 \geq 1 ] \leq \Pr [ \Ll \geq \rho v ] \leq  \exp(-\lambda \rho^2 v^2)
$.
The result follows directly from Theorem \ref{th:lbexp}.
\end{proof}

\begin{corollary}
If $r <  \rho v$, then
$
\E [\tau] \geq \dfrac{2}{3\sqrt{\lambda} v} (1/(\theta r) - 1).
$\end{corollary}
\begin{proof}
If $r < \rho v$, then $a_{i_0}$ can only infect the agents moving in the same direction as him: its connection time with the agents moving in the opposite direction is upper bounded by $r/v < \rho$.
Let $N_c$ be the number of agents that $a_{i_0}$ connects to while being on $s_0$. Since each agent in $s_0$ can be moving in any of the directions with a probability $1/2$, $N_c$ is dominated by a random Poisson variable with parameter $2 \theta r / 2 = \theta r$, which means that 
\begin{align*}
    \Pr [N_c \geq 1]
    &= \int_0^{+\infty}  \Pr [N_c \geq 1 \colon L_{\lambda}^{(0)} = \ell ] \fl(\ell) d\ell\\
    &\leq \int_0^{+\infty} 1 - e^{-\theta r} \fl(\ell) d\ell
    = 1 - e^{-\theta r} \leq \theta r.
\end{align*}
Since $C_0 \leq N_c$, Theorem \ref{th:lbexp} gives the desired result.
\end{proof}
\begin{corollary}\label{cor:lm/th>>1}
We have
$
\E [\tau] \geq (\sqrt{\lambda}/\theta - 4/3)/(2\sqrt{\lambda} v).
$\end{corollary}
\begin{proof}
We can easily prove that the number $N$ of agents that were on $s_0$ at some time instant when $a_{i_0}$ was on it too follows a Poisson distribution with parameter $2\theta L_{\lambda}^{(0)}$. In fact, when $a_{i_0}$ enters $s_0$, there are $N_1$ agents on the street, and by the time it reaches the end of the street, since all the agents have the same speed, all of these will have left it and $N_2$ new agents will have come. $N_1$ and $N_2$ both follow a Poisson distribution with parameter $\theta L_{\lambda}^{(0)}$ and $N = N_1 + N_2$. Finally, given that $C_0 \leq N$, we have 
\begin{align*}
    \Pr [C_0 \geq 1] 
    & \leq \Pr [N \geq 1]
      = \int_0^{+\infty}  \Pr [N \geq 1 \colon L_{\lambda,0} = l ] \fl(l) dl \\
    & = \int_0^{+\infty} (1 - e^{-2\theta l}) \fl(l) dl \\
    & \leq \int_0^{+\infty} \frac{2\theta l}{\sqrt{\lambda}} f_L(l) dl
    = \frac{2\theta}{\sqrt{\lambda}} \E[ L_1 ] = \frac{4 \theta}{3\sqrt{\lambda}}.
\end{align*}
Applying Theorem \ref{th:lbexp} concludes the proof.
\end{proof}
Using Theorem \ref{th:lbprob} in these three cases, we can also find lower bounds for $\tau$ that hold with high probability.
\section{Simulation results}\label{sec:simres}
\label{PE}
This section discusses simulations that were performed to analyze malware propagation in D2D, to benchmark the mathematical study made in Section~\ref{TS} and to show how the various parameters accelerate or slow down propagation.  
Our ABM was built based on Mesa \cite{mesa}, which is a \textcolor{black}{ very suitable} python framework for ABMs that we have extended to generate and visualize street system environments.

\subsection{Evaluation indicators}
We present some indicators that allow us to analyze malware propagation. They should be independent of the dimensions of the map, since we theoretically want to study propagation on an infinite plan. 

\begin{definition}[\textbf{Propagation speed}]\label{def:propspeed}
Propagation speed is the velocity of malware spread in space. It is defined by
\[
\propspeed:= \limsup\limits_{u \rightarrow + \infty} u\E[ 1/\tau_u],
\]
with $\tau_u$ the time when the infection reaches the distance $u$ from the initial infection point
\[
\tau_u := \inf\{ t \geq 0 \colon \exists a_j \in \I(t) : \| X_j(t) - X_{I_0}(0) \| \geq u \}.
\]
\end{definition}
To study the system behaviour, we will set a value of $u$ large enough and observe $\propspeed_u$ considering that it approximates sufficiently the asymptotic values.
We remind that we denote by $\I(t)$ the set of infected agents at time $t$, and that we call $a_{I_0}$ the only initially infected agent, and $X_{I_0}$ its position at time $0$. $a_{I_0}$ is always chosen close to the center of the map.

\begin{definition}[\textbf{Infection rate}]\label{def:rinf}
The infection rate is the rate of infected agents in the region reached by the virus
\[
\rinf := \limsup\limits_{u \rightarrow + \infty} \dfrac{|\I(\tau_u)|}{| \{ X_j(\tau_u) \colon a_j \in \A\} \cap B(X_{I_0}(0), u) |},
\]
where $B(X_{I_0}(0), u)$ is the open ball of center $X_{I_0}(0)$ and radius $u$, and $\tau_u$ is as in Definition \ref{def:propspeed}.
\end{definition}

Note that $ | \{ X_j(\tau_u) \} \cap B(X_{I_0}(0), u) |$ is simply the number of agents inside $B(X_{I_0}(0), u)$ at time $\tau_u$.

$\propspeed$ and $\rinf$ are defined as limits, let $\propspeed_u$ and $\rinf_u$ be the expressions in Definitions \ref{def:propspeed} and \ref{def:rinf} that converge to them respectively. Since the plots we will make require running lots of simulations, and thus take a very long time to be constructed, we were not able to make them with different values of $u$ and study the convergence of $\propspeed_u$, $\rinf_u$ to $\propspeed$, $\rinf$. Since we are mostly interested in the behavior of the system and not really in the exact values of the propagation speed and the infection rate, it is enough to set a large enough value for $u$, consider that $\propspeed \approx \propspeed_u$ and $\rinf \approx \rinf_u$, and interpret the results. 

\subsection{Simulation results} \label{sec:default_simul}
For all simulations, unless otherwise stated, parameters are set by default as follows: 
($u = 3.5 km, H = 10km, \lambda = 50 km^{-2}, \theta = 3 km^{-1}, v= 5 km/h, \rho = 20s, r = 200m$). where $H$ is the side length of the square surface containing the map.
We assume $dt = 0.9 \rho$. Each value in the diagrams we will present later is the average over 20 simulations with the same set of parameters. In the diagrams where $\lambda$ does not vary, we use the same 20 maps for all the points.

\subsubsection{The threshold $\sqrt{\lambda}\rho v$}
The critical regimes seen in Section~\ref{TS} are relevant and confirm the intuitive expectations one may have for the virus propagation. However, the most remarkable result concerns the regime $\sqrt{\lambda} \rho v \gg 1$, because the lower bound found for $\E[\tau]$ grows with a speed of $x \mapsto \exp(x^2)/x$ in the quantity $\sqrt{\lambda} \rho v$, we can thus expect to observe a rather tight threshold at the level of which the propagation is no longer possible.
To have meaningful results, we will vary $\lambda$ from $10$ to $200$ and the speed of the agents from $1$ to $90$, and the other parameters will be set by default as in Sections~\ref{sec:default_simul}. However, when $\lambda$ is very large, the number of agents $\E[ |\A| ] = 2 \sqrt{\lambda} H^2 \theta$ will be also large since, even if it is only proportional to $\sqrt{\lambda}$, the multiplicative constant is large. To keep a reasonable number of agents, we use maps with side-length $H_{\lambda} := 20\lambda^{-1/4}$ for each value of $\lambda$, and the stopping propagation radius $u_{\lambda} := 0.45 \times H_\lambda$ to have $H > 2u$. This will guarantee that the expected number of agents is $\E[\A] = 2400$ ($\theta = 3$), and the side-lengths will vary from $\approx 11.24$ to $\approx 5.32 km$.
\begin{figure}
\centering
\begin{tikzpicture}
\node (img)  {\includegraphics[width=.9\linewidth]{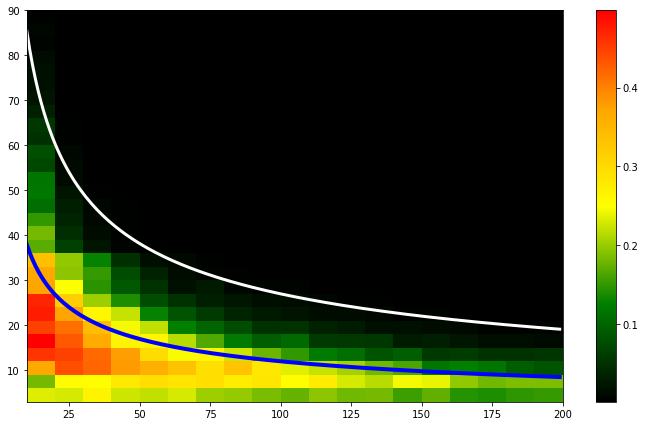}};
\node[below=of img, node distance=0cm, yshift=1.1cm] {$\lambda$};
\node[left=of img, node distance=0cm, rotate=0, anchor=center,yshift=0cm, xshift=1cm] {$v$};
\end{tikzpicture}
\vspace{-.4cm}
\caption{Infection rate $\rinf$}
\label{fig:nowkpvt_rinf}
\vspace{-.5cm}
\end{figure}%
\begin{figure}
\centering
\begin{tikzpicture}
\node (img)  {\includegraphics[width=.9\linewidth]{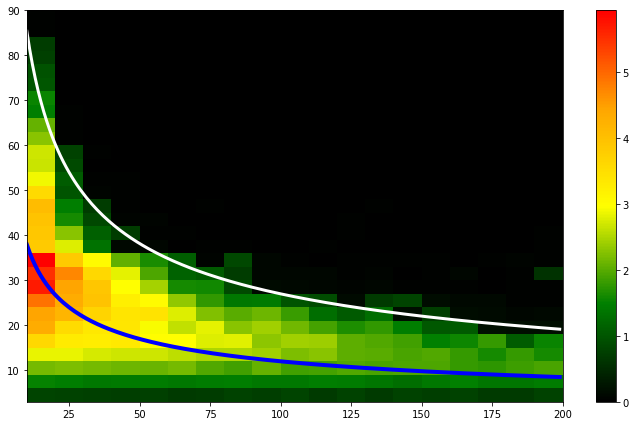}};
\node[below=of img, node distance=0cm, yshift=1.1cm] {$\lambda$};
\node[left=of img, node distance=0cm, rotate=0, anchor=center,yshift=0cm, xshift=1cm] {$v$};
\end{tikzpicture}
\vspace{-.4cm}
\caption{Propagation speed $\propspeed$ in $(km/h)$}
\label{fig:nowkpvt_propspd}
\vspace{-.5cm}
\end{figure} 

We observe in Figures~\ref{fig:nowkpvt_rinf} and  \ref{fig:nowkpvt_propspd} that the rate of infection and the speed of propagation both cancel out above a certain threshold curve, having a shape of type \textcolor{black}{ $v(\lambda)= c/(\rho\sqrt{\lambda})$, as indicated in blue (for $c=2/3$) and white (for $c=3/2$)}. This confirms the hypothesis of the exponential lower bound of $\E[\tau]$, although it is obtained with a simplified mathematical model. It seems however that this threshold is sharper for $\rinf$ than for
$\propspeed$. The reason why we have such a threshold is that the distribution of the edges lengths in a PVT makes it very rare to have edges much larger than the mean edge length $\E[ L_\lambda ]
=2/(3\sqrt{\lambda})$ (see Corollary \ref{cor:moments}), and therefore, when there is no edge larger than $\rho v$, the virus cannot propagate since connection require agents to be on the same street.

With respect to Figure~\ref{fig:nowkpvt_propspd}, we see that the virus can hardly propagate if $\sqrt{\lambda} \rho v \geq 3/2$. A surprising remark is that the maximum infection rate is always not far below the curve $\sqrt{\lambda} \rho v = 2/3$, while the maximum propagation speed seems to be achieved exactly at the points verifying this equation. 
We also observe a lower threshold value of the speed: the virus hardly propagates for $v = 3$, but as soon as $v = 6$, we see a remarkable jump in the values of $\rinf$ and $\propspeed$. It is to be expected to have a weaker propagation for the small values of the speed because in the limit $v=0$ the virus can propagate at most in the street where it was initially placed. 

The third observation is that the virus propagation becomes slower as $\lambda$ becomes larger. The reason is that, as predicted by the simplified model in Section~\ref{TS}, when $\sqrt{\lambda}$ becomes much larger than $\theta$, we have too many streets compared to the number of agents, and therefore $a_{I_0}$ will only meet a few agents.
\subsubsection{How is the propagation speed impacted by $\theta$ and $v$?}
\textcolor{black}{ The propagation speed of the virus is certainly a function of all the parameters of our model. 
However, the distance $r$ is given by the technology and cannot be changed, and the intensity of streets $\lambda$ is known for a given city. 
Now, for a given malware, we want to see the influence of the intensity and speed of users on the propagation speed and the infection rate.}
In fact, agents that move fast enough but not too fast, i.e., not to have $\sqrt{\lambda} \rho v \geq 3/2$, will rapidly carry the virus to the other edges and facilitate its spreading. Also, when agents' intensity is important, there will be always agents on these streets that will get infected and carry the virus further.
\begin{figure}
  \centering
\begin{tikzpicture}
\node (img)  {\includegraphics[width=.9\linewidth]{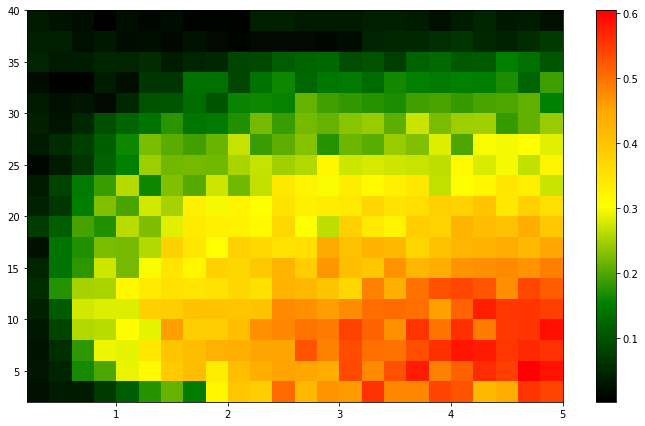}};
\node[below=of img, node distance=0cm, yshift=1.1cm] {$\theta$};
\node[left=of img, node distance=0cm, rotate=0, anchor=center,yshift=0cm, xshift=1cm] {$v$};
\end{tikzpicture}
  \vspace{-.4cm}
  \caption{Infection rate $\rinf$}
  \label{fig:th_v_rinf}
  \vspace{-.5cm}
\end{figure}%
\begin{figure}
  \centering
  \begin{tikzpicture}
\node (img)  {\includegraphics[width=.9\linewidth]{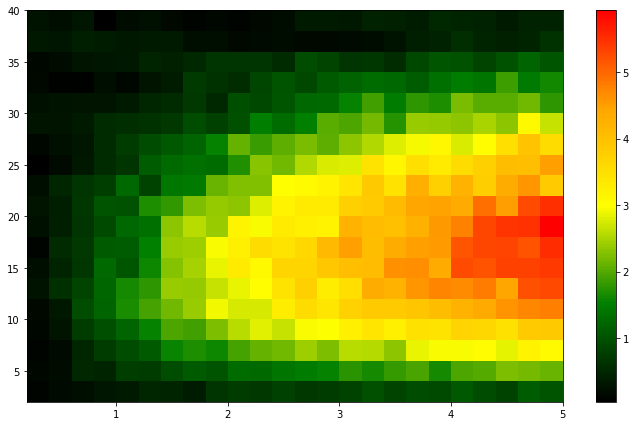}};
\node[below=of img, node distance=0cm, yshift=1.1cm] {$\theta$};
\node[left=of img, node distance=0cm, rotate=0, anchor=center,yshift=0cm, xshift=1cm] {$v$};
\end{tikzpicture}
\vspace{-.4cm}
  \caption{Propagation speed $\propspeed$ in $(km/h)$}
  \label{fig:th_v_propspeed}
  \vspace{-.5cm}
\end{figure}

Considering Figure~\ref{fig:th_v_propspeed}, we see that the propagation speed and the infection rate show different behaviors. Indeed, although both are increasing in $\theta$, $\rinf$ is maximal for $v$ around $7-10km/h$, while $\propspeed$ is maximal for $v$ around $15-20km/h$. Moreover, the high values of the propagation speed are more concentrated while those of $\rinf$ seem to be more spread out. Also, for every $\theta$, there is clearly an increase and then a decrease of $\propspeed$ when we increase $v$, going from $\approx 0km/h$ to the maximal value and then returning to $0km/h$. But the value $\rinf$ does not change a lot in the first range of values of $v$. This means that, when agents are slow, they will stay sufficiently long on every street and therefore, once an infected agent reaches a street, it will infect many agents being on it too. Propagation speed is nevertheless slow because agents take a lot of time before exiting each street and carrying the virus to the next one.  
This correlation between $\rinf$ and $\propspeed$ confirms the need to study these two quantities together.

Returning to the results of Figure~\ref{fig:th_v_rinf}, the value of $v$ for which $\sqrt{\lambda}\rho v = 2/3$ is $v_0 \approx 16.97$. Thus, we have again that $\rinf$ is maximal in the region below the level line $\sqrt{\lambda}\rho v = 2/3$, and $\propspeed$ is maximal exactly in its close neighborhood. This property would therefore be true even when varying $\theta$.
For larger values of $v$, we expect that the virus will not propagate anymore because the streets are not long enough, and we already see the beginning of this behavior. However, we notice that the speed at which the propagation weakens depends on $\theta$: the higher the intensity of the agents, the higher the speed needed to weaken the virus propagation, which is to be expected since the increase of $\theta$ favors the propagation of the virus. Moreover, for small values of $\theta$, the propagation never takes place whatever the value of $v$ because the agents are few and do not establish enough connections ($\theta$ is below the percolation threshold).

\section{Conclusion and future work}
\label{conc}
This paper presents a novel ABM for analyzing malware propagation dynamics in D2D networks. This approach, traditionally applied for complex systems, allows us to obtain relevant and surprising findings about malware propagation in D2D, which demonstrate also the effectiveness for such dynamical communication networks. Notably, malware propagation was not possible above a first threshold ($\sqrt{\lambda} \rho v > 3/2$) and was maximal around a second threshold ($\sqrt{\lambda} \rho v = 2/3$), which corresponds to having an average length of streets equal to the distance traveled by an agent during the time $\rho$ (needed for infection transmission). This shows
the importance of street system characteristics, which has been traditionally neglected when studying malware propagation in D2D. We believe that the ABM approach has a great potential for studying malware spread in D2D communication networks.
Besides generalizations such as adding attributes for the street widths, devices out of the street system or sojourn times, as future work, we aim to model and simulate countermeasure policies for reversing malware attacks. 

\bibliographystyle{IEEEtran}
\bibliography{geostoch}
\end{document}